\documentclass{moriond}

\def\be{\begin{equation}}
\def\ee{\end{equation}}
\def\bea{\begin{eqnarray}}
\def\eea{\end{eqnarray}}

\newcommand{\Photo}{}

\begin{document}


\vspace*{4cm}
\title{RECENT RESULTS ON ANISOTROPIC FLOW AND RELATED PHENOMENA IN ALICE}

\author{ ANTE BILANDZIC (FOR ALICE COLLABORATION) }

\address{Physik Department E62, Technische Universit\"{a}t M\"{u}nchen,\\ 85748 Garching, Germany}

\maketitle\abstracts{The exploration of properties of an extreme state of matter, the Quark--Gluon Plasma, has broken new ground with the recent Run 2 operation of the Large Hadron Collider with heavy-ion collisions at the highest energy to date. With the heavy-ion data taken at the end of 2015, the ALICE Collaboration has made the first observation of anisotropic flow of charged particles and related phenomena in lead-–lead collisions at the record breaking energy of 5.02 TeV per nucleon pair. The Run 2 results come after the proton-lead collisions, which provided a lot of unexpected results obtained with two- and multi-particle correlation techniques. In these proceedings, a brief overview of these results will be shown. We will discuss how they further enlighten the properties of matter produced in ultrarelativistic nuclear collisions. We indicate the possibility that, to leading order, the striking universality of flow results obtained with correlation techniques in pp, p--A and A--A collisions might have purely mathematical origin, and that physical conclusions therefore could be drawn only from the subleading orders.}


\section{Introduction}

Properties of an extreme state of nuclear matter, the Quark--Gluon Plasma (QGP), produced in ultra-relativistic heavy-ion collisions can be studied by measuring the anisotropic flow phenomenon with correlation techniques~\cite{Ollitrault:1992bk,Wang:1991qh,Borghini:2001vi}. Anisotropic flow is sensitive both to the initial geometry of the volume containing the produced nuclear matter and to its transport properties and equation of state. Anisotropic pressure gradients developed in an interacting medium are a sufficient condition for anisotropic flow to develop, which turns into the observable anisotropic distribution of produced particles in heavy-ion collisions. Such anisotropic distributions can be parameterized with the azimuthal angle $\varphi$ and are conveniently quantified with the Fourier series~\cite{Voloshin:1994mz}:
\begin{equation}
f(\varphi) = \frac{1}{2\pi}\left[1+2\sum_{n=1}^{\infty}v_n\cos[n(\varphi-\Psi_n)]\right]\,.
\label{eq:Fourier}
\end{equation}
In this sense, flow harmonics $v_n$ and symmetry planes $\Psi_n$ quantify the anisotropic flow phenomenon, and both degrees of freedom can be related to the QGP properties. Using the orthogonality relations of trigonometric functions, from Eq.~(\ref{eq:Fourier}) it follows that $v_n = \left<\cos[n(\varphi\!-\!\Psi_n)]\right>$, where angular brackets denote an average over all particles in an event. However, this relation cannot be used in practice to measure flow harmonics $v_n$ due to in experiment unknown symmetry planes $\Psi_n$. Instead, estimates for $v_n$ can be obtained using correlation techniques involving two or more particles, for which the only required input are the azimuthal angles $\varphi$ of reconstructed particles in a heavy-ion collision. 

Anisotropic flow is a collective phenomenon typically engaging all produced particles. This unique feature can be used to discriminate flow correlations from correlations stemming from other physical phenomena, which predominantly occur only among few particles (so called nonflow correlations). When only anisotropic flow is present, any apparent correlation among produced particles can be attributed solely to the existence of common anisotropic pressure gradients, which develop as a response of the strongly interacting system to the initial spatial anisotropies characteristic for non-central heavy-ion collisions. Written mathematically, in this case any joint multi-variate probability density function (p.d.f.) of $n$ azimuthal angles $\varphi_1,\ldots,\varphi_n$ fully factorizes:
\begin{equation}
f(\varphi_1,\ldots,\varphi_n) = f_{\varphi_{1}}(\varphi_{1})\cdots f_{\varphi_{n}}(\varphi_n)\,,
\label{eq:factorization}
\end{equation}
where on the right hand side is the product of the normalized marginalized p.d.f.'s, $f_{\varphi_i}(\varphi_i)$, $1\leq i\leq n$, each of which is the same~\cite{Danielewicz:1983we} and is given by Eq.~(\ref{eq:Fourier}). Due to above factorization, any azimuthal correlation can be related to $v_n$ and $\Psi_n$ degrees of freedom, introduced in Eq.~(\ref{eq:Fourier}), and therefore each one in principle can provide an independent estimate for them. When factorization in Eq.~(\ref{eq:factorization}) breaks down due to presence of nonflow, more reliable estimates for $v_n$ and $\Psi_n$ can be obtained with multi-particle cumulants, which by construction are less sensitive to nonflow effects as the order of correlator increases~\cite{Borghini:2001vi}. 


\section{Anisotropic flow in small systems?}

The recent flow measurements with multi-particle cumulants in the collisions of light and heavy nuclei, like p--Pb at LHC~\cite{Aad:2013fja,Abelev:2014mda,Khachatryan:2015waa} or p--Au, d--Au and ${}^3$He--Au at RHIC~\cite{Adamczyk:2015xjc,Adare:2015ctn}, have flared a lot of discussions, both among theorists and experimentalists. Since to leading order these measurements resemble the features observed in the heavy-ion collisions~(see Fig.~\ref{fig:Fig1}, left and middle panels), which are attributed to collective anisotropic flow, it is very tempting to interpret them the same way for smaller systems. This interpretation is challenged by the outcome of Monte Carlo studies for $e^+e^-$ systems~\cite{Loizides:2016tew} in which neither the QGP existence nor collective effects are expected, where to leading order multi-particle cumulants exhibit yet again the similar universal trends, both for $v_2$ and $v_3$ harmonics. 

In the next section we provide a new possible explanation for such a universal leading order behaviour seen everywhere.


\section{Correlation techniques in a nutshell}

Correlation techniques are applicable in the anisotropic flow analysis only if all effects of self-correlations are exactly removed. This can be achieved by expressing all azimuthal correlations analytically in terms of $Q$-vectors~\cite{Danielewicz:1985hn}, evaluated for different harmonics~\cite{Bilandzic:2010jr,Bilandzic:2013kga}. The $Q$-vector for harmonic $n$ is a complex number which is defined for a set of $M$ particles as: 
\begin{equation}
Q_{n} \equiv \sum_{k=1}^{M}\,e^{in\varphi_k} \,,
\label{eq:Qvector}
\end{equation}
where $\varphi_k$ labels the azimuthal angle of the $k$th particle. Each multi-particle azimuthal correlation can be expressed as a combination of the $Q$-vectors. This implies that all of their statistical properties can be determined from the statistical properties of the $Q$-vectors. For any random observable, the statistical properties are completely determined by its p.d.f., or equivalently by its characteristic function, $\phi(k)$, which by definition is the inverse Fourier transform of p.d.f. Recently, the analytic expressions for the characteristic functions of real and imaginary parts of $Q$-vectors were derived for the most general case of anisotropic flow for $M$ particles~\cite{Bilandzic:2014qga}:
\begin{equation}
\phi_{{\rm Re}\,Q_m}(k) = \left[J_0(k) + 2\sum_{p=1}^\infty\,(-1)^p\left[c_{2p\cdot m}J_{2p}(k)
-  i c_{(2p-1)\cdot m} J_{2p-1}(k)\right]\right]^M\,,
\label{eq:CharacteristicFunction_Re_M-particle}
\end{equation}
and
\begin{equation}
\phi_{{\rm Im}\,Q_m}(k) = \left[J_0(k) + 2\sum_{p=1}^\infty\,\left[c_{2p\cdot m}J_{2p}(k)
+  i s_{(2p-1)\cdot m} J_{2p-1}(k)\right]\right]^M\,.
\label{eq:CharacteristicFunction_Im_M-particle}
\end{equation}
Harmonics $c_n$ and $s_n$ in the above expressions originate from the alternative parametrization~\footnote{Two parametrization can be trivially related by using relations $v_n=\sqrt{c_n^2+s_n^2}$ and $\Psi_n=(1/n)\arctan\frac{s_n}{c_n}$.} of initial single-particle Fourier-like p.d.f. in Eq.~(\ref{eq:Fourier}). Therefore, the characteristic functions of real and imaginary parts of the $Q$-vector are solely given in terms of Bessel functions of the first kind. This has a remarkable consequence for the statistical properties of $Q$-vectors, because both of the above characteristic functions are dominated by the $J_0(k)$ term, since this is the only term which is not weighted with the initial Fourier harmonics $c_n$ and $s_n$ (which are typically smaller than 0.1 in magnitude), and since all $J_0(k)$, $J_1(k)$, $J_2(k)$, etc., are comparable in magnitude. This leads us to conclude that, since the dominant $J_0(k)$ term is always present (even for the case of random walk), to leading order we will always see universal trends in the distributions of $Q$-vectors, i.e. their distributions exhibit a purely mathematical attractor. Since all multi-particle azimuthal correlations can be expressed analytically in terms of $Q$-vectors, we conjecture that the striking universality of flow results obtained with correlation techniques in pp, p--A and A--A collisions can be attributed solely to the fact that they exhibit a purely mathematical attractor as well. The analogous derivation of characteristic functions for multi-particle azimuthal correlations, which at the present is out of reach, will either prove or disprove this conjecture.~\footnote{A plausible, but clearly not formal, argument for the existence of such an attractor can be drawn immediately from the central limit theorem, since by definition multi-particle azimuthal correlations are sums of random observables, for which due to central limit theorem the attractor always exists, and it is Gaussian.} If proved correct, any flow analysis relying on multi-particle correlation techniques would need to draw physical conclusions only from sub-leading orders, and the observed leading order universalities in a vastly different colliding systems per se would have no physical meaning. 


\section{First flow results from Run 2}

\begin{figure}
\begin{minipage}{0.66\linewidth}
\centerline{\includegraphics[width=1.0\linewidth]{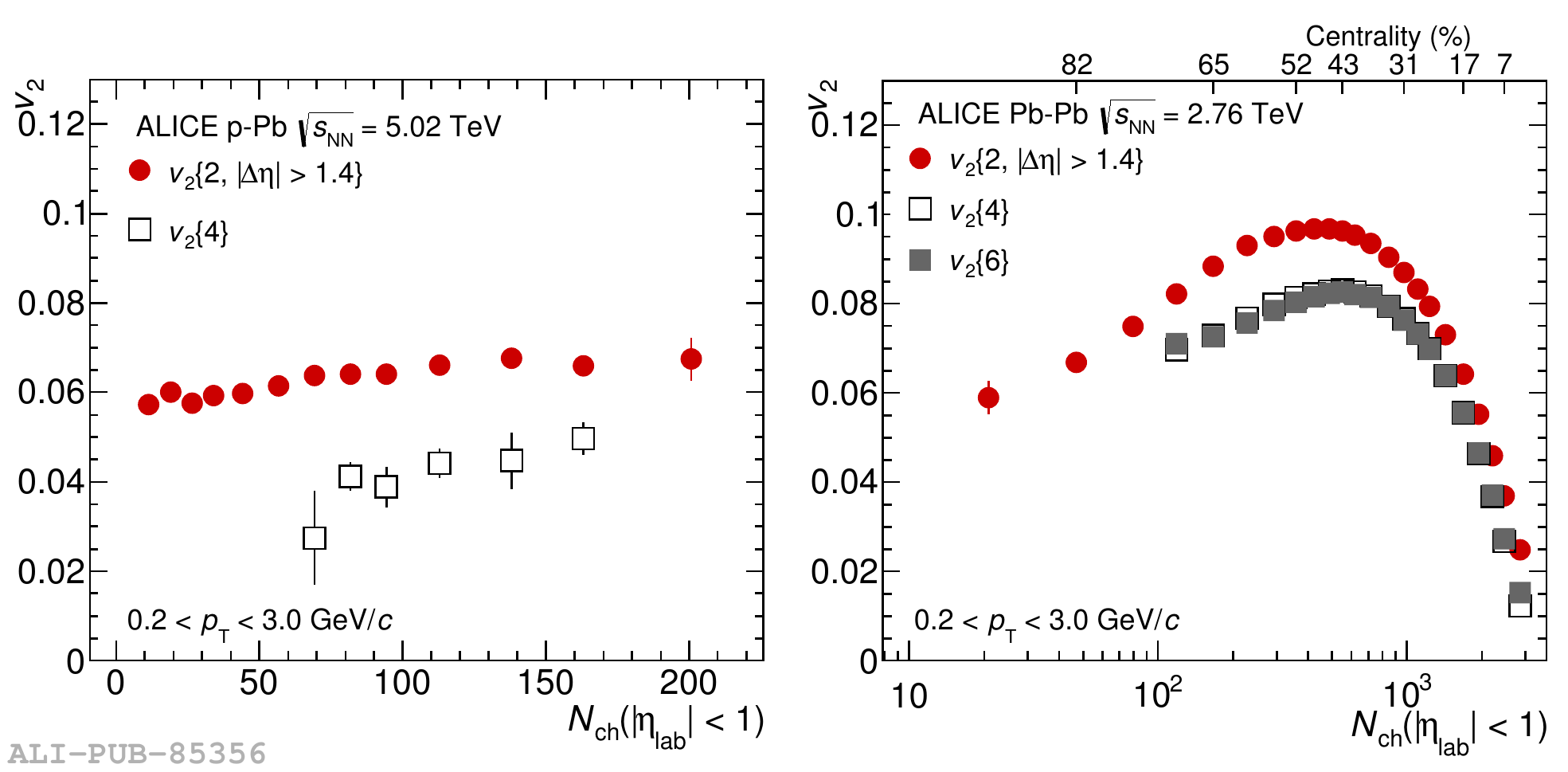}}
\end{minipage}
\hfill
\begin{minipage}{0.33\linewidth}
\centerline{\includegraphics[width=1.0\linewidth]{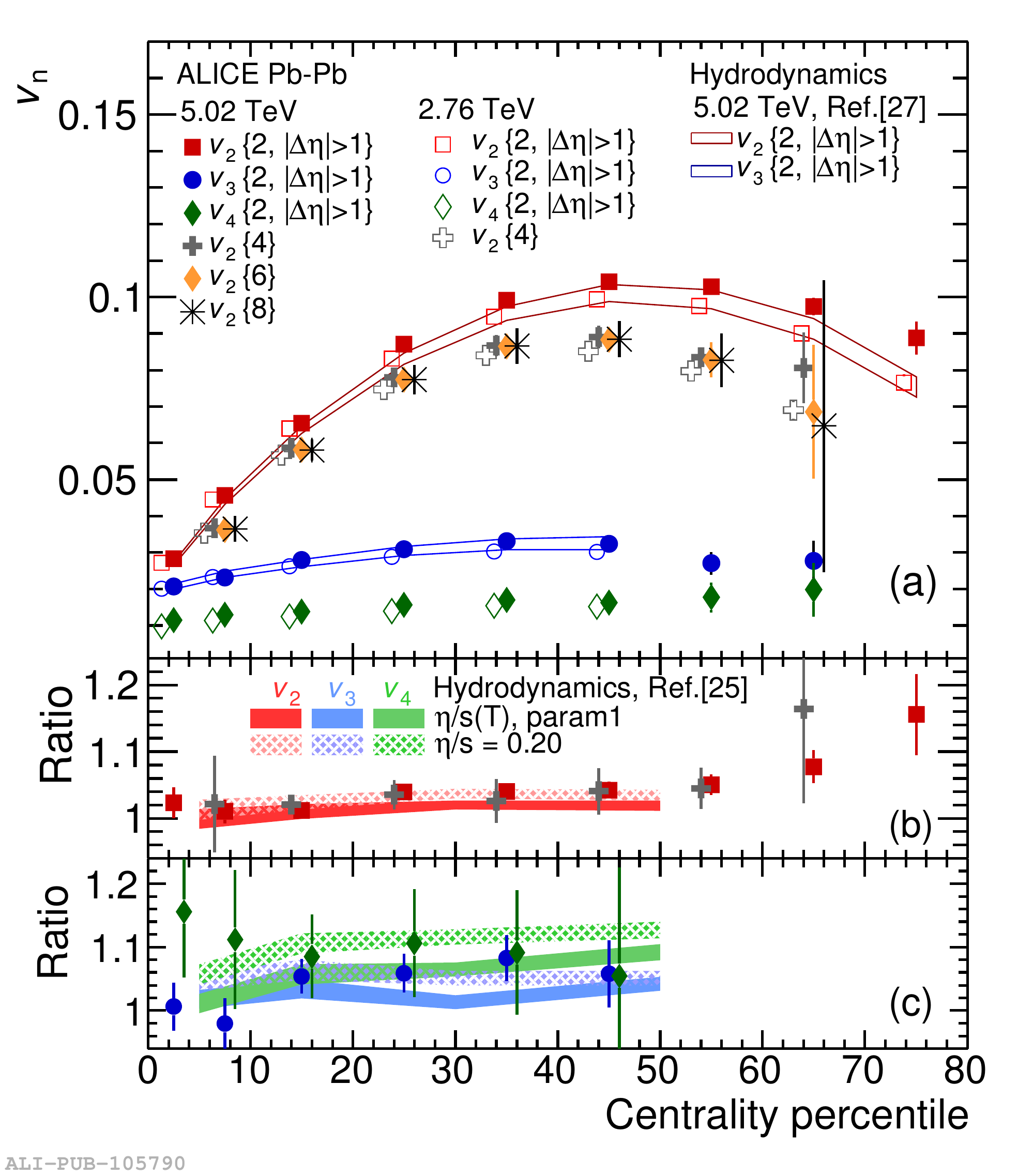}}
\end{minipage}
\caption[]{(color online) Multiplicity dependence of $v_2$ estimated with two- and multi-particle correlation techniques, in p--Pb (left panel) and Pb--Pb (middle panel), at Run~1 LHC energies~\cite{Abelev:2014mda}. Centrality dependence of flow harmonics in Pb--Pb collisions at Run 1 and Run 2 LHC energies, compared to theoretical models (right panel), see~\cite{Adam:2016izf} and references therein.}
\label{fig:Fig1}
\end{figure}

Finally, we report the first results of $v_2$, $v_3$ and $v_4$ of charged particles in Pb--Pb collisions at a center-of-mass energy per nucleon pair of $\sqrt{s_{_{\rm NN}}}$ = 5.02 TeV~\cite{Adam:2016izf}. Theoretical predictions for the dependence of anisotropic flow on collision energy can be found elsewhere~\cite{Auvinen:2013ira,Noronha-Hostler:2015uye,Shen:2012vn,Niemi:2015qia,Niemi:2015voa}. The data used were recorded with the ALICE detector in November 2015 in Run 2 at the LHC. A sample of 140 k Pb--Pb collisions passed the selection criteria. The measurements are performed in the central pseudorapidity region $|\eta| < 0.8$ and for the transverse momentum range $0.2 < p_{\rm T} < 5$ GeV/$c$. Compared to Run 1 results from Pb--Pb collisions at $\sqrt{s_{_{\rm NN}}}$ = 2.76 TeV, the anisotropic flow coefficients $v_2$, $v_3$ and $v_4$ are found to increase (see Fig.~\ref{fig:Fig1}, right panel) on average by $(3.0\pm 0.6)\%$, $(4.3\pm 1.4)\%$ and $(10.2\pm 3.8)\%$, respectively, in the centrality range $0\!-–\!50$\%, which is found to be compatible with hydrodynamic model calculations~\cite{Noronha-Hostler:2015uye,Niemi:2015voa}. The detailed theoretical study of various parameterizations of the temperature dependence of shear viscosity to entropy density ratio was performed in~\cite{Niemi:2015voa}, out of which the Run 2 flow results seem to favor the constant value for shear viscosity to entropy density ratio. The transverse momentum dependence of anisotropic flow does not change appreciably between the two LHC energies, which indicates that the increase in integrated flow coefficients can be attributed mostly to an increase in average transverse momentum between Run 1 and Run 2 LHC energies~\cite{Adam:2016izf}.
%
%

%
%




\end{document}